\documentclass{svproc}
\usepackage{url}

\usepackage{amsmath,amsfonts,amssymb}
\usepackage{stmaryrd}
\usepackage{graphicx}
\usepackage[hidelinks]{hyperref}
\usepackage[utf8]{inputenc}
\usepackage[small]{caption}
\usepackage{cleveref}
\usepackage[font=scriptsize,labelfont=scriptsize]{subcaption}
\usepackage[normalem]{ulem}
\usepackage{multicol}
\usepackage{multirow}
\usepackage{soul}
\usepackage{color}

\begin{document}
\mainmatter              
%
\title{The Benefits of Interaction Constraints \\ in Distributed Autonomous Systems}
%
\titlerunning{The Benefits of Interaction Constraints}  
%
\author{Michael Crosscombe\inst{1} \and Jonathan Lawry\inst{2}}
%
\authorrunning{M. Crosscombe and J. Lawry} 
%
%
\institute{Graduate School of Arts and Sciences, University of Tokyo, Japan\\
\email{cross@sacral.c.u-tokyo.ac.jp}
\and
Department of Engineering Mathematics, University of Bristol, United Kingdom\\
\email{j.lawry@bristol.ac.uk}}


\maketitle              

\begin{abstract}
    The design of distributed autonomous systems often omits consideration of the underlying network dynamics. Recent works in multi-agent systems and swarm robotics alike have highlighted the impact that the interactions between agents have on the collective behaviours exhibited by the system. In this paper, we seek to highlight the role that the underlying interaction network plays in determining the performance of the collective behaviour of a system, comparing its impact with that of the physical network. We contextualise this by defining a collective learning problem in which agents must reach a consensus about their environment in the presence of noisy information. We show that the physical connectivity of the agents plays a less important role than when an interaction network of limited connectivity is imposed on the system to constrain agent communication. Constraining agent interactions in this way drastically improves the performance of the system in a collective learning context. Additionally, we provide further evidence for the idea that `less is more' when it comes to propagating information in distributed autonomous systems for the purpose of collective learning.
\end{abstract}
\section{Introduction}
\label{sec:introduction}

During the development of autonomous systems, it is only natural that we focus our attention on the aspects of a system which are responsible for producing the desired behaviours.
In distributed autonomous systems the focus is often on how the interactions between individual agents can lead to effective macro-level collective behaviours.
What may be oversimplified or even forgotten, however, are the underlying networks which underpin those interactions.
A typical assumption to make is that the agent's \emph{interaction network} is merely a product of the spatial properties of the system, i.e.\ agents interact randomly with other agents based only on whether or not they are within the required distance for interaction to occur; often this takes the form of a communication range in swarm robotics using platforms such as the Kilobots with infrared transceivers that are limited to a physical communication radius of $10$ cm~\cite{Rubenstein2014,Valentini2017}. However, in this paper we argue that we ought to consider {\em which} agents should be allowed to interact, rather than merely considering {\em how} those agents interact.


A common modelling assumption in distributed multi-agent and multi-robot systems is the ``well-stirred system'' assumption: In a well-stirred system each agent is equally likely to encounter any other agent in the system and, therefore, each interaction is regarded as an independent event~\cite{parker2009cooperative}. This equates to a system in which the interaction network of the agents is totally connected and edges are selected at random to simulate the random interactions of different pairs of agents. However, this assumption does not consider that intentional constraints placed on agent interactions may yield better performance than if the interaction network is assumed to be totally connected and interaction depends only upon the physical proximity of the agents. More recently, \cite{talamali2021less} found that ``less is more'' when it comes to communication in a robot swarm and that, by reducing the physical communication radius of the robots (or the density of the swarm), the system-level performance improves as a result of there being less communication overall.
While this is the most recent study to highlight the problem of too much communication, \cite[Sec. 3.2]{hamann2018superlinear} previously demonstrated a similar problem in which agents became ``too cooperative''. This general problem can be traced further back to \cite{lazer2007network} in which the authors examined the effect of network structure on system-level performance.
A general survey of recent literature that considers the connectivity of a multi-agent/robot system conducting collective exploration is provided in \cite{Kwa2022}.

In previous work we have studied the impact of applying constraints to the interaction network in a non-physical multi-agent system within a collective learning setting~\cite{crosscombe2021impact}. We showed that the ``less is more'' hypothesis holds true for collective learning when the underlying interaction network's topology is a small-world network and that interaction networks with low connectivity lead to optimal performance for collective learning. However, this study was limited to studying interaction networks in abstract simulations, without the ability to consider the physical network layer. Therefore, in this work we propose to study both the physical and interaction networks in combination, to determine which layer of connectivity is more important for performance in the development of distributed autonomous systems. 

An outline of the remainder of this paper is as follows: We begin by describing collective learning as the context of our investigation, before defining a three-valued model for collective learning. We present results for a totally-connected interaction network but where the physical network connectivity is varied, so as to understand the extent to which the physical network has an impact on system performance. Then, we introduce constraints over the interaction network by restricting its topology to that of a regular lattice network, before studying system performance with respect to different levels of connectivity. Finally, we conclude our findings with some discussions and highlight our main contributions before detailing directions for future work.


\section{Collective Learning in Multi-Agent Systems}
\label{sec:collective-learning}

Social learning broadly describes the observed phenomena in animals where an individual's learning is influenced by observing and interacting with other individuals or their products~\cite{heyes1994social}. An example of this is the use of pheromones to facilitate teaching in the tandem-running ant {\it Temnothorax albipennis}~\cite{franks2006teaching}. This style of learning forms the basis of many distributed autonomous systems which, by design, rely on the interactions of individual agents to produce behaviours more complex than could be demonstrated by any one individual.

In the last decade, distributed decision-making has been identified as a foundational problem in swarm systems due to the need for large groups of agents to cooperate on a shared task by functioning as a collective~\cite{brambilla2013,schranz2020}. In swarm systems, the ability for agents to make decisions in a \emph{distributed} manner is crucial for their real-world deployment, as robot swarms are appropriate platforms for deployment in difficult terrain and hazardous environments, such as for the detection of wildfires~\cite{innocente2019}. Due to their large scale, swarm systems often cannot be supervised at the individual level and must instead rely on human supervision at the swarm level~\cite{hogg2022}, while making its own decisions and taking actions autonomously, often by reaching a consensus~\cite{valentini2017book} or through quorum sensing~\cite{bechon2012}.
Much of the literature on collective decision-making in robot swarms centres on the best-of-$n$ problem; a bio-inspired problem in which the swarm must decide which is the best option of $n$ possible alternatives~\cite{parker2009cooperative}. However, this class of decision-making is limited to scenarios in which each option is associated with a quality value. Instead, we focus on a state-of-the-world problem whereby agents must learn the current state of the world, in order to make informed decisions about their assigned task.

In this paper we adopt the term {\em collective learning} to describe social learning in a population of artificial agents attempting to reach a consensus about the state of their environment~\cite{crosscombe2021collective}. 
In this context, collective learning is used specifically to describe the combination of two distinct processes: {\em direct learning} based on evidence obtained directly from an agent's environment, and {\em indirect learning} in the form of an agent fusing its belief with the belief of another agent. The effects of combining these two processes has been studied from an epistemological perspective by \cite{douven2011truth} who argue that communication between agents acts both to propagate information through the population, and to correct for errors that occur whilst gathering evidence. The connection between these two processes has been further investigated more recently in \cite{crosscombe2016,douven2019optimizing}.


\section{A Three-Valued Model for Collective Learning}
\label{sec:model}

We consider a collective learning problem in which a distributed system of autonomous agents attempts to learn the true state of their environment.
Suppose that the environment can be described by a set of $n$ propositional variables $\mathcal{P} = \{p_1, \dots, p_n\}$ with each relating to a discrete location, then an agent's belief about the environment is an allocation of truth values to each of the propositional variables $b : \mathcal{P} \rightarrow \{0, \frac{1}{2}, 1\}^{n}$. In our model we adopt a three-valued propositional logic with the truth values $0$ and $1$ denoting false and true, respectively, whilst the third truth-value $\frac{1}{2}$ is used to denote {\em unknown} and allows agents to express uncertainty about the environment. For notational convenience, we can instead represent an agent's belief $b$ by the $n$-tuple $B = \langle B(p_1), \ldots, B(p_n) \rangle$. By doing so, an agent can express uncertainty about a location in the environment by assigning the truth value $\frac{1}{2}$ to any of the propositional variables in $\mathcal{P}$. As an example, let us suppose that an environment can be described by $n = 2$ propositional variables, then the belief $B = \langle 0, 1 \rangle$ would represent that the agent believes $p_1$ to be false, while $p_2$ is true. Now let us suppose another agent holds the belief $B^\prime = \langle \frac{1}{2}, \frac{1}{2} \rangle$, then the agent is expressing uncertainty about both of the propositions $p_1$ and $p_2$, thereby indicating that the agent is totally uncertain about the environment.

We now define the two processes of direct and indirect learning in the context of our proposed three-valued model and describe how they combine to produce collective learning at the macro level in distributed autonomous systems.

\paragraph{Evidential updating.} The learning process by which an agent obtains evidence {\em directly} from its environment, e.g., through onboard sensory capabilities, and updates its current belief to reflect this evidence is referred to as {\em evidential updating}. In our model, an agent first decides which proposition to investigate by selecting a single proposition {\em at random} from those propositions about which the agent is uncertain, i.e.\  where $B(p_i) = \frac{1}{2}$. Then, having selected a proposition $p_i$ to investigate, the agent travels to the corresponding location in the environment where it gathers evidence. Evidence takes the form of an assertion about the truth value of the chosen proposition $p_i$ such that $E = \langle \frac{1}{2}, \ldots, S^*(p_i), \ldots, \frac{1}{2} \rangle$, where $S^* : \mathcal{P} \rightarrow \{0,1\}$ denotes the ground truth -- often referred to as the {\em true state of the world} -- and that the truth value of each proposition is certainly true or false.
Upon obtaining evidence $E$ the agent then updates its belief to $B|E$ as follows:
\begin{align}
    \begin{split}
        B|E &= \langle B(p_1) \odot E(p_1), \dots, B(p_n) \odot E(p_n) \rangle\\
        &= B \odot E
    \end{split}
    \label{eq:evidence}
\end{align}
Where $\odot$ is a binary operator defined on $\{0, \frac{1}{2}, 1\}$ given in \Cref{tab:fusion}. We will elaborate on the reasoning behind our choice of fusion operator when we introduce the belief fusion process but, for now, notice that the operator works for evidential updating by preserving {\em certainty} over {\em uncertainty} and that updating in this manner does not, therefore, alter the truth value of any proposition $p_j \in \mathcal{P}$ where $p_j \neq p_i$ because $E(p_j) = \frac{1}{2}$. An agent repeatedly gathers evidence about uncertain propositions until it has become totally certain about the environment, at which point it ceases to look for evidence.

In real-world environments the act of gathering evidence is often subject to noise, stemming either from the use of noisy sensors -- such as thermal sensors used by UAVs~\cite{deVries} -- or from the environment itself. In this model evidence shall therefore take the following form:
\begin{gather}\label{eq:noisy-evidence}
    E(p_i) = 
    \begin{cases}
        ~ S^*(p_i) &: ~\text{ with probability }1 - \epsilon \\
        ~ 1 -  S^*(p_i) &: ~\text{ with probability } \epsilon
    \end{cases}
\end{gather}
where $\epsilon \in [0,0.5]$ is a noise parameter denoting the probability that the evidence is erroneous.

\paragraph{Belief fusion.} In {\em collective} learning the agents do not exist in isolation. Instead, individual agents can benefit from being able to interact with, and learn from, other agents in a shared environment.
To this end, we now describe the {\em indirect} learning process by which pairs of agents fuse their beliefs in order to improve the system's ability to learn the state of the world. 

\begin{table}[t]
    \centering
    \setlength{\tabcolsep}{6pt}
    \def\arraystretch{1.4}
    \begin{tabular}{ccccc}
        & \multicolumn{4}{l}{\hspace*{0.65cm}$B^\prime(p_i)$} \\ \cline{2-5} 
        \multicolumn{1}{l|}{\multirow{4}{*}{$B(p_i)$}} &
        \multicolumn{1}{l|}{$\odot$} &
        \multicolumn{1}{l|}{$\boldsymbol{0}$} &
        \multicolumn{1}{l|}{$\boldsymbol{\frac{1}{2}}$} &
        \multicolumn{1}{l|}{$\boldsymbol{1}$} \\ \cline{2-5} 
        \multicolumn{1}{l|}{} &
        \multicolumn{1}{l|}{$\boldsymbol{0}$} &
        \multicolumn{1}{l|}{0} &
        \multicolumn{1}{l|}{0} &
        \multicolumn{1}{l|}{$\frac{1}{2}$} \\ \cline{2-5}
        \multicolumn{1}{l|}{} &
        \multicolumn{1}{l|}{$\boldsymbol{\frac{1}{2}}$} &
        \multicolumn{1}{l|}{0}&
        \multicolumn{1}{l|}{$\frac{1}{2}$} &
        \multicolumn{1}{l|}{1} \\ \cline{2-5}
        \multicolumn{1}{l|}{}&
        \multicolumn{1}{l|}{$\boldsymbol{1}$} &
        \multicolumn{1}{l|}{$\frac{1}{2}$} &
        \multicolumn{1}{l|}{1} &
        \multicolumn{1}{l|}{1} \\ \cline{2-5} 
    \end{tabular}
    \vskip 0.25cm
    \caption{The fusion operator $\odot$ applied to beliefs $B$ and $B^\prime$.}
    \label{tab:fusion}
\end{table}

The belief fusion process provides two crucial benefits: Firstly, agents are able to propagate information they obtain directly from the environment through the system by fusing their beliefs with those of other agents. We propose to combine beliefs using the three-valued fusion operator in \Cref{tab:fusion} which is applied element-wise to all of the propositional variables $p_i \in \mathcal{P}$. Given two beliefs $B$ and $B^\prime$, corresponding to the beliefs of two agents, the fused belief is then given by:
\begin{align}
    B \odot B^\prime = \langle B(p_1) \odot B^\prime(p_1), \dots, B(p_n) \odot B^\prime(p_n) \rangle
    \label{eq:fusion}
\end{align}

Secondly, the belief fusion operator helps to correct for errors that occur when the evidence obtained by the agents is subject to noise.
This error-correcting effect arises when two agents have gathered evidence about the same proposition $p_i$, independently of one another, and encounter a disagreement. When a pair of agents differ in their belief about the truth value of $p_i$, then the belief fusion process leads them both to become {\em uncertain} with regards to $p_i$.
As an example, suppose that two agents hold the beliefs $B_1(p_i) = 1$ and $B_2(p_i) = 0$, then upon the agents fusing their beliefs such that $B_1 \odot B_2(p_i) = \frac{1}{2}$, both agents will attempt to seek additional evidence about proposition $p_i$, either directly from the environment or indirectly via fusion with other agents, having become uncertain about the truth value of $p_i$. 
Notice that the belief fusion process can lead to agents becoming {\em less certain} when the fusing agents disagree about the truth value of a given proposition, while evidential updating only leads to agents becoming {\em more certain}.

\paragraph{Measuring performance.} In the context of collective learning agents are attempting to learn a representation of the true state of the world underlying their environment. As such, we can measure the accuracy of the {\em average belief} of the population with respect to the true state of the world in the following way: Given a population of $m$ agents, the normalised difference between each agent's belief $B$ and the true state of the world $S^*$, averaged across the population, is the average error of the system;
\begin{gather}
    \frac{1}{m} \frac{1}{n} \sum_{j = 1}^m \sum_{i=1}^n \left| B_j(p_i) - S^*(p_i) \right|
    \label{eq:average-error}
\end{gather}
This shall be our primary metric of performance.

\section{Simulation Environment}
\label{sec:simulation}

\begin{figure*}[ht!]
    \begin{center}
        \begin{subfigure}{.25\textwidth}
            \includegraphics[width=1\textwidth]{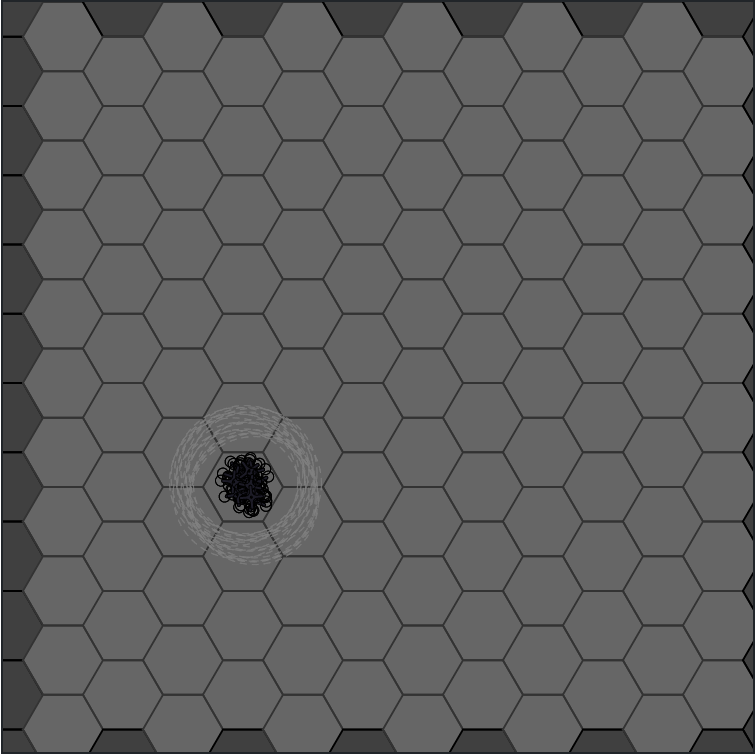}
            \caption{At initialisation.}
            \label{fig:sim-a}
        \end{subfigure}
        \hspace{1em}
        \begin{subfigure}{.25\textwidth}
            \includegraphics[width=1\textwidth]{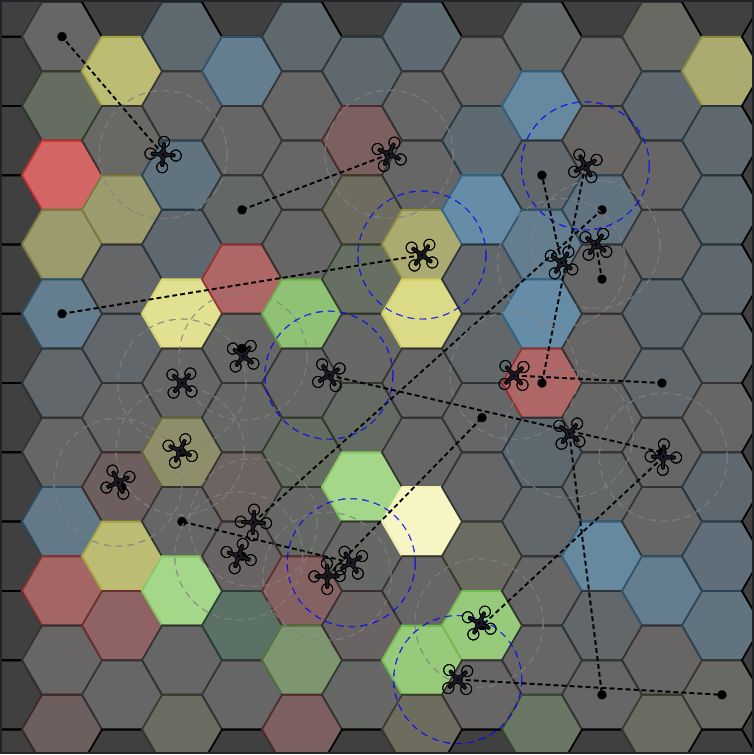}
            \caption{In progress.}
            \label{fig:sim-b}
        \end{subfigure}
        \hspace{1em}
        \begin{subfigure}{.25\textwidth}
            \includegraphics[width=1\textwidth]{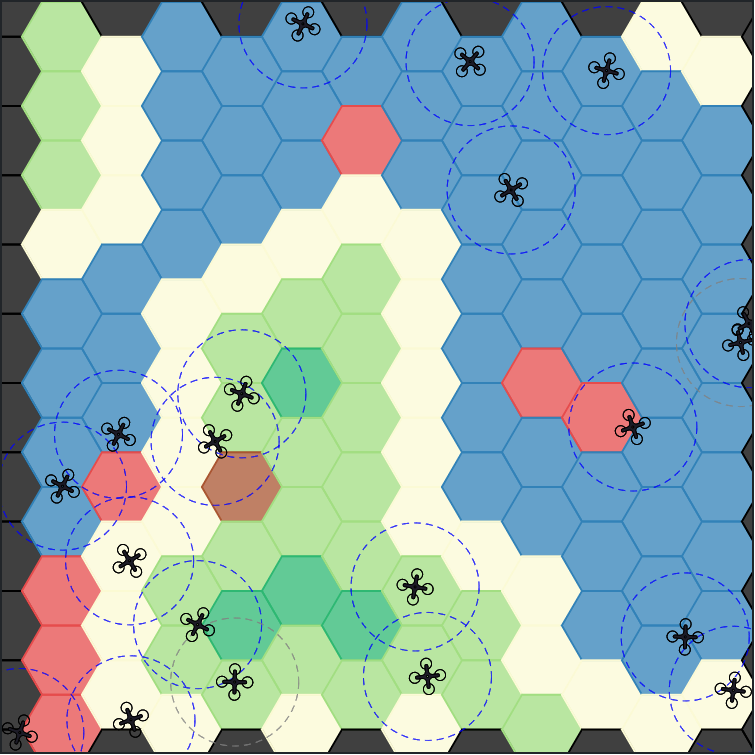}
            \caption{At convergence.}
            \label{fig:sim-c}
        \end{subfigure}
        \vskip 0.25cm
        \caption{
            Simulation environment depicting a noisy collective learning scenario in which $m=20$ agents attempt to learn about $n=126$ locations when the environment noise $\epsilon = 0.2$. Each agent has a communication radius $C_r = 20$.
        }
        \label{fig:simulation-env}
    \end{center}
\end{figure*}

We now introduce the physical simulation environment used to study collective learning in a multi-agent system.
In \Cref{fig:simulation-env} the simulation environment is depicted with $20$ agents and $126$ hexagonal locations for the agents to investigate\footnote{We found that $20$ agents were sufficient to complete the task with performance similar to that of $50$ agents.}.
Agents are represented by black Unmanned Aerial Vehicles (UAVs) with either a grey or a blue dashed circle surrounding them, indicating their physical communication radius $C_r$. If two agents are within each other's physical radii of communication then communication between those two agents is \emph{possible}. If the radius is blue, then an agent is said to be in a communicating state, whilst grey indicates that it is not communicating. At initialisation (\cref{fig:sim-a}), the environment is greyed out to represent the collective lack of information that the system holds about its environment. During the collective learning process (\cref{fig:sim-b}), agents begin to build an internal representation of the state of the world, which can be averaged across the system to give us a general view of performance. To show this, the colours of the hexagons increase in intensity representing the collective certainty about each location as contained in the belief of the average agent. The different colours are used to represent ocean (blue), beach (yellow), and low and high terrain (light and dark green), but are purely visual in this context. The brown hexagon is the initial launch location from which all agents begin exploring. In the case that the system collectively adopts the incorrect truth value for a given proposition, the hexagon at its associated location is coloured red, with its intensity still representing the average certainty with which the agents have (incorrectly) determined its truth value. At the end of the simulation (\cref{fig:sim-c}) the agents have reached a consensus about the state of the world and the map is in full view, albeit with some errors depicted by red hexagons.


\section{The Networks Underlying Agent Interactions}
\label{sec:underlying-networks}

\begin{figure*}[t!]
    \begin{center}
        \begin{subfigure}{.19\textwidth}
            \includegraphics[width=1\textwidth]{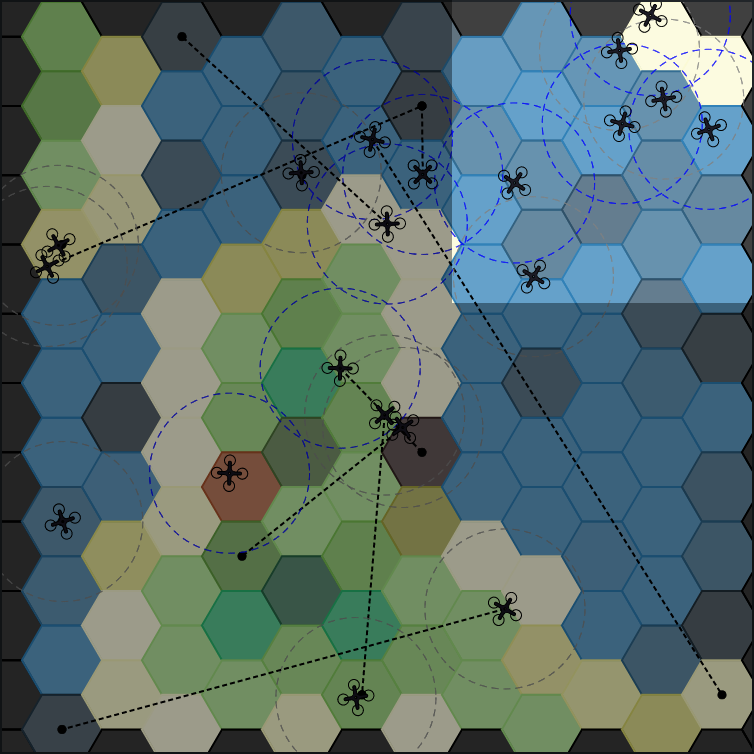}
            \caption{}
        \end{subfigure}
        \begin{subfigure}{.19\textwidth}
            \includegraphics[width=1\textwidth]{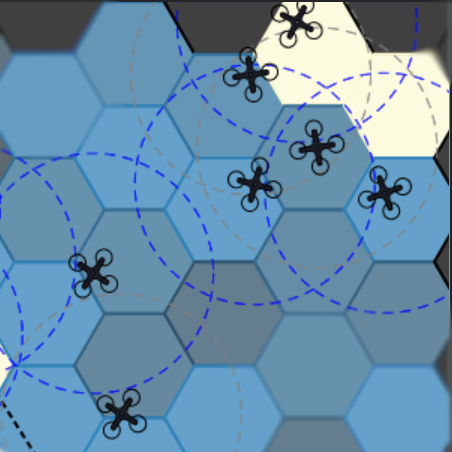}
            \caption{}
        \end{subfigure}
        \begin{subfigure}{.19\textwidth}
            \includegraphics[width=1\textwidth]{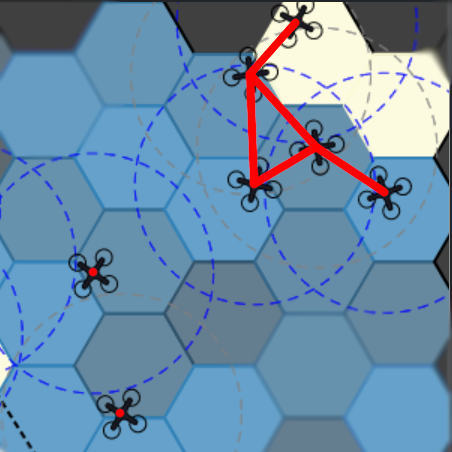}
            \caption{}
        \end{subfigure}
        \begin{subfigure}{.19\textwidth}
            \includegraphics[width=1\textwidth]{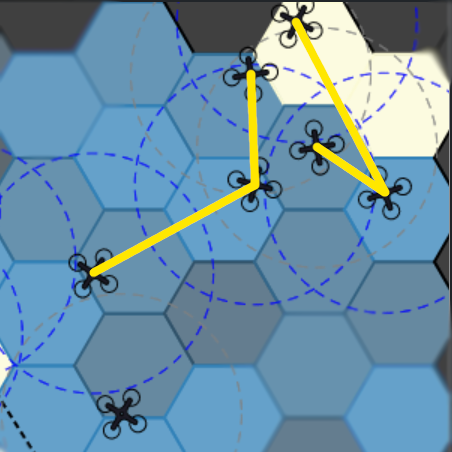}
            \caption{}
        \end{subfigure}
        \begin{subfigure}{.19\textwidth}
            \includegraphics[width=1\textwidth]{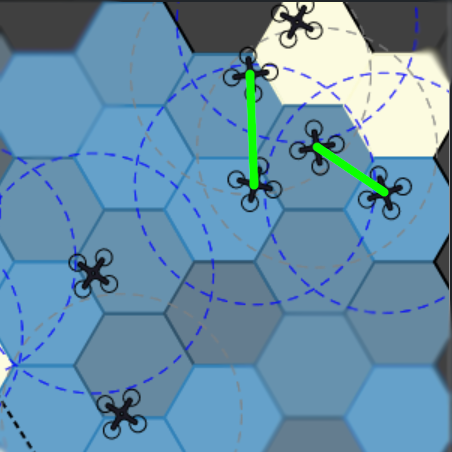}
            \caption{}
        \end{subfigure}
        \vskip 0.25cm
        \caption{A snapshot of the simulation during an experiment. In this model two different types of networks are formed which govern the agent interactions in the system.
        (a) A snapshot of the simulation environment during runtime. (b) A zoomed-in view of 7 agents with a system-wide communication radius $C_r = 20$. (c) A representation of the physical network that forms from the agents' communication radii. (d) A representation of the pre-assigned interaction network between the visible agents. (e) The resulting communication links that are available according to edges which are present in both the physical and interaction networks.}
        \label{fig:simulation-networks}
    \end{center}
\end{figure*}

When we consider the design of distributed autonomous systems (e.g. multi-agent systems or robot swarms), the system-level performance is driven solely by the interactions of its constituent agents with their environment and with each other. The consequence of this fact is that we focus our effort on designing the rules and processes which define {\em how} the agents interact, but often we neglect to consider {\em which} agents ought to interact. In the past, we have often assumed that it is beneficial for agents to be able to interact freely with any other agent in the system, but recent works have shown this to be false~\cite{crosscombe2021impact,talamali2021less}. To understand the extent to which agents interact in our simulation, we now distinguish between the physical network and the interaction network as they occur in our simulation environment.

In \Cref{fig:simulation-networks} we show a snapshot of the simulation during runtime in which we overlay visual representations of the different networks which are formed during a single time step $t$. Assuming each agent acts as a node in both a physical network and an interaction network, then the {\bf red} edges connect pairs of agents which are {\em physically capable} of communicating, based on the agents being within their communication range with radius $C_r$. Edges in the physical network are added or removed during each time step $t$ of the simulation according to which pairs of agents are within communication range of one another.
The {\bf yellow} edges, meanwhile, depict the connectivity of the interaction network as determined at the start of each experiment. Unlike in the physical network, the pairs of agents which are {\em allowed} to interact are immutable, such that no new edges are added, nor existing edges removed after the network is initialised.
The {\bf green} edges depict which edges exist in both networks at a given time step $t$. Each edge then corresponds to a pair of agents which are both able and allowed to interact with one another.

An interaction network can be constructed in whatever manner leads to optimal system performance, provided that the agents remain {\em capable} of interacting physically in the environment during their task. The default interaction network in many multi-agent and multi-robot systems is a totally-connected network. In this paper we seek to constrain interactions by reducing the connectivity of the network whilst improving system performance. To do this we shall use regular lattice networks, having discovered in previous work that such a topology is more optimal even than small-world networks in this setting~\cite{crosscombe2021impact,watts1998strogatz}.


\section{Multi-Agent Simulation Experiments}
\label{sec:experiment}

For each experimental condition, the system proceeds in the following way: We initialise $m = 20$ agents at the same starting location and with total uncertainty about the $n = 126$ locations, i.e.\ each agent begins with the belief $B(p_i) = \frac{1}{2}$ for $p_i \in \mathcal{P}$. Every agent is also initialised with the same communication radius $C_r \in \{20, \ldots, 100\}$ and a communication frequency $C_f \in [0, 1]$.

After initialisation, each agent selects a proposition $p_i$ at random about which they are uncertain and begins travelling to the associated location in search of evidence about $p_i$. When they arrive at the location the agent will observe their environment and obtain evidence $E$ according to \cref{eq:noisy-evidence}, before updating its belief by \cref{eq:evidence}.
The agent will then select another uncertain proposition $p_i$ to investigate and begin to travel in the direction of the new location. At this moment, the agent will also enter a communicating state with probability $C_f$. If the agent does not enter the communicating state then it continues to its destination to obtain evidence and the process repeats. If the agent does enter the communicating state, however, then the agent begins broadcasting its belief to any agents that come within its communication radii $C_r$ as it travels to its destination. The agent remains in the communicating state until it encounters one or more other communicating agents within communication range. 
Upon receiving the belief(s) of the other agent(s), the agent will cease communicating and update its belief $B$ by selecting another belief $B\prime$ at random before adopting the fused belief $B \odot B\prime$ by \cref{eq:fusion}. If the agent remains uncertain about the location to which it is travelling, then the agent continues to that location as normal. However if the agent, having fused its belief, becomes certain about the location during travel then it simply repeats the process of searching for new evidence by first selecting a new uncertain location from its belief.
Finally, if instead the agent holds a totally certain belief, such that $B(p_i) \in \{0,1\}$ for all $p_i \in \mathcal{P}$, then the agent repeatedly enters a communicating state in an attempt to reach a consensus with the rest of the system.

Each experiment runs for a maximum of $30,000$ time steps or until the population has reached a unanimous consensus about their environment. We repeat each experiment $50$ times for each parameter combination $C_r$ and $C_f$ and for three different noise levels $\epsilon \in \{0, 0.1, 0.3\}$ representing noise-free, low- and moderate-noise scenarios, respectively. The results are then averages across the $50$ runs. We now present results for simulation experiments where the underlying interaction network is totally-connected, in keeping with the ``well-stirred system'' assumption, in which we study how changes to the connectivity of the physical network affect the macro-level dynamics of the system.


\subsection{Convergence Results for Physical Networks}
\label{ssec:results}

\begin{figure*}[t!]
    \begin{center}
        \begin{subfigure}{1\textwidth}
            \includegraphics[width=1\textwidth]{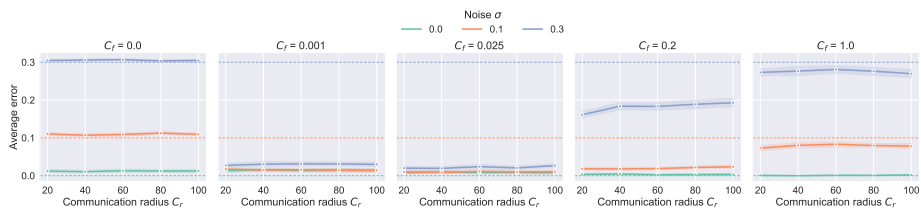}
        \end{subfigure}
        \vskip 0.25cm
        \caption{Average error at steady state against communication radius $C_r$ for increasing communication frequency $C_f = [0, 1]$. Each dotted line shows the noise $\epsilon \in \{0, 0.1, 0.3\}$. The shaded regions show a $95\%$ confidence interval.
        }
        \label{fig:complete-error-radius}
    \end{center}
\end{figure*}


In this section we show convergence results for a static totally-connected interaction network and focus our attention on the dynamic physical network for which the connectivity changes based on the chosen communication radius $C_r$. In all figures, the shaded regions show a $95\%$ confidence interval.

In \Cref{fig:complete-error-radius} we show the average error of the population (\cref{eq:average-error}) at steady state against communication radius $C_r$. Each solid line shows the results for different noise $\epsilon$, while the dashed lines of the same colour show the noise $\epsilon$. Each plot within this figure represents a different communication frequency $C_f$ increasing from $0$ (no communication) to $1$ (unrestricted communication).
For $C_f = 0$ we present a special case in which agents do not communicate with one another (referred to as ``asocial learning'' by \cite{heyes1994social}). As a result of this total absence of belief fusion, the agents rely solely upon evidential updating to inform their beliefs and, in the case where the evidential updating process is noisy, the agents are unable to learn an accurate representation of their environment. The result is the system reaching an average error around $\epsilon$ for all communication radii $C_r$. Therefore, $\epsilon$ serves as a simple benchmark of performance and any reduction in the average error to below $\epsilon$ demonstrated improved performance.

As we increase $C_f$ and agents are able to fuse their beliefs, we see an immediate and significant reduction in the average error for $C_f \geq 0.001$ such that for all noisy scenarios (i.e.\ for $\epsilon = 0.1$ and $0.3$) the average error is reduced to below the level of noise in each scenario. In the noise-free scenario (green line) the populations converge to an average error above $0$ when the communication frequency is too low (i.e.\ for $C_f < 0.1$, suggesting that while the population has converged to a highly accurate {\em average belief}, the population does not reach a consensus across the system. Here we see that the best performance is measured for a communication frequency $C_f = 0.025$, but increasing $C_f$ further results in an increased average error and, therefore, worse performance in the system.

Perhaps surprisingly, increasing the communication radius $C_r$ of the system, and thereby the connectivity of the physical network over which agents communicate, has little effect on the macro-level performance as measured by the average error of the system at steady state. The lines in \Cref{fig:complete-error-radius} remain relatively flat across different communication radii $C_r$, with the performance of the system being almost entirely determined by the communication frequency $C_f$. This suggests that the level of connectivity of the physical network has very little impact on the performance of the system for the purpose of collective learning when agents can communicate freely provided they are in range.


\section{Constraining the Interaction Network}
\label{sec:networks}

\begin{figure}[ht!]
    \begin{center}
        \begin{subfigure}{.2\textwidth}
            \begin{center}
                \includegraphics[width=0.65\textwidth]{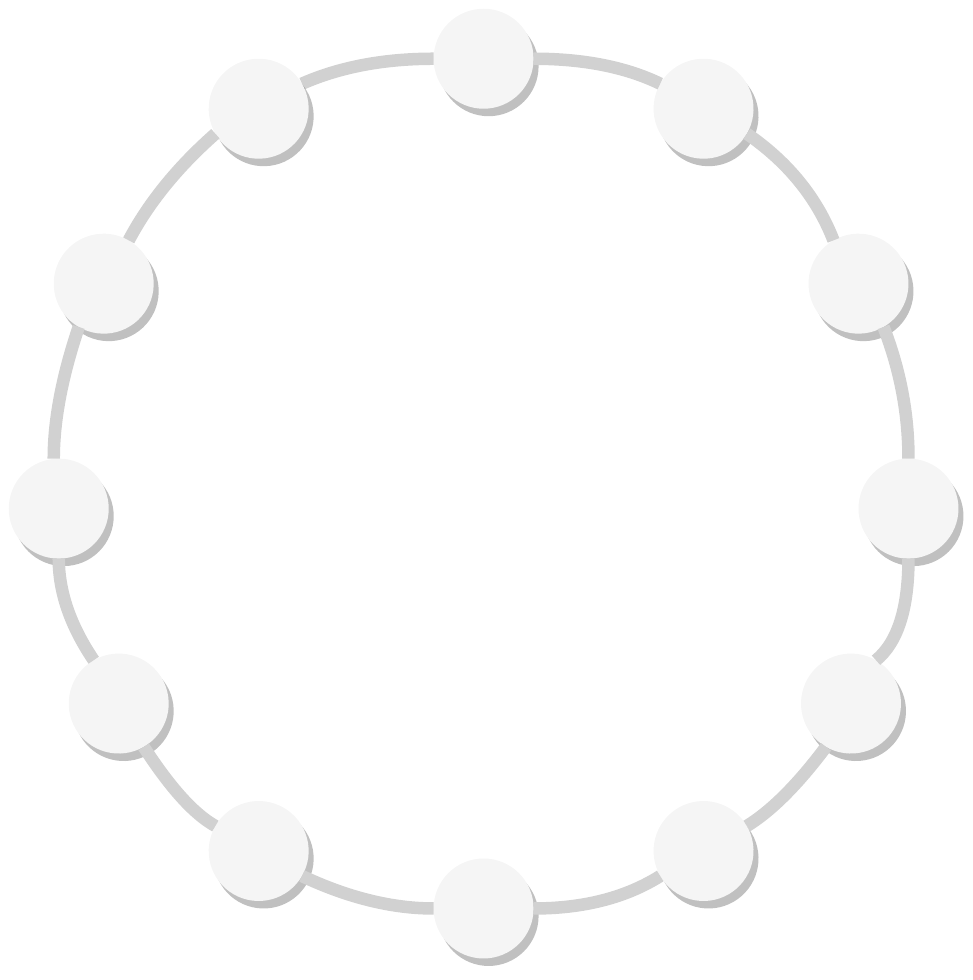}
                \caption{}
                \label{diag:swn-k2}
            \end{center}
        \end{subfigure}
        \hspace{2em}
        \begin{subfigure}{.2\textwidth}
            \begin{center}
                \includegraphics[width=0.65\textwidth]{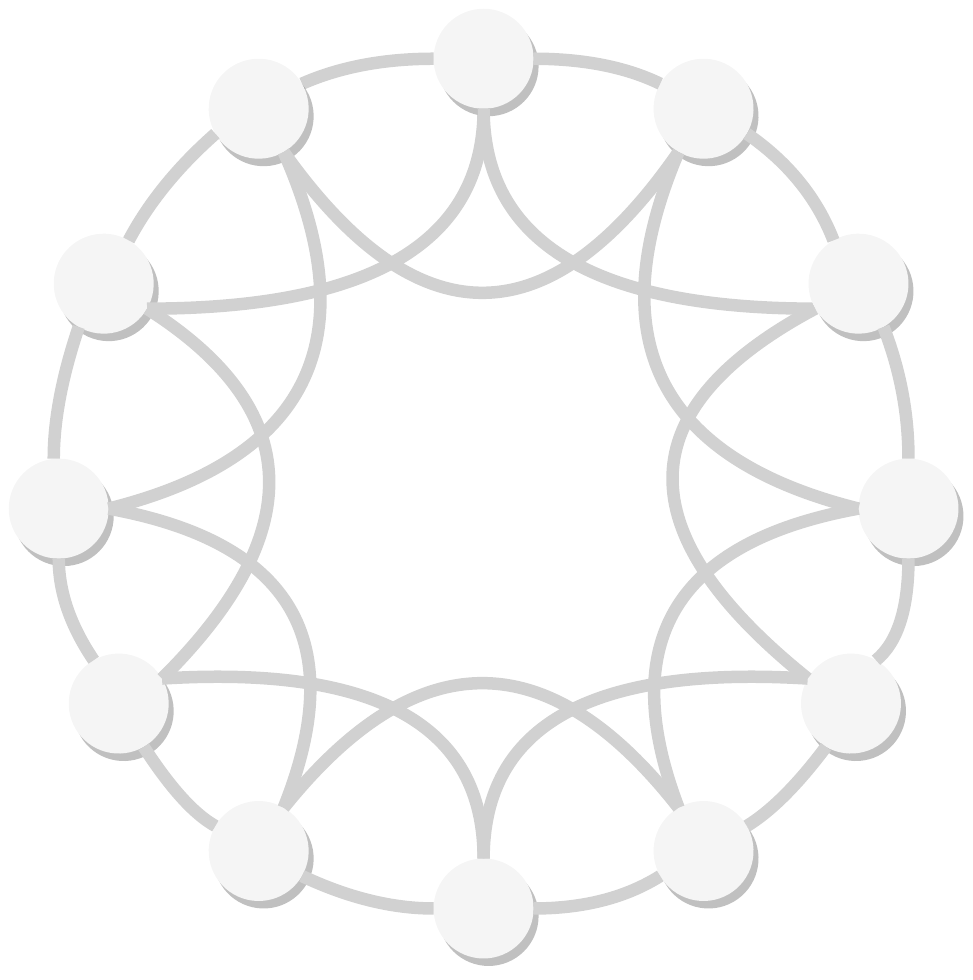}
                \caption{}
                \label{diag:swn-k4}
            \end{center}
        \end{subfigure}
        \hspace{2em}
        \begin{subfigure}{.2\textwidth}
            \begin{center}
                \includegraphics[width=0.65\textwidth]{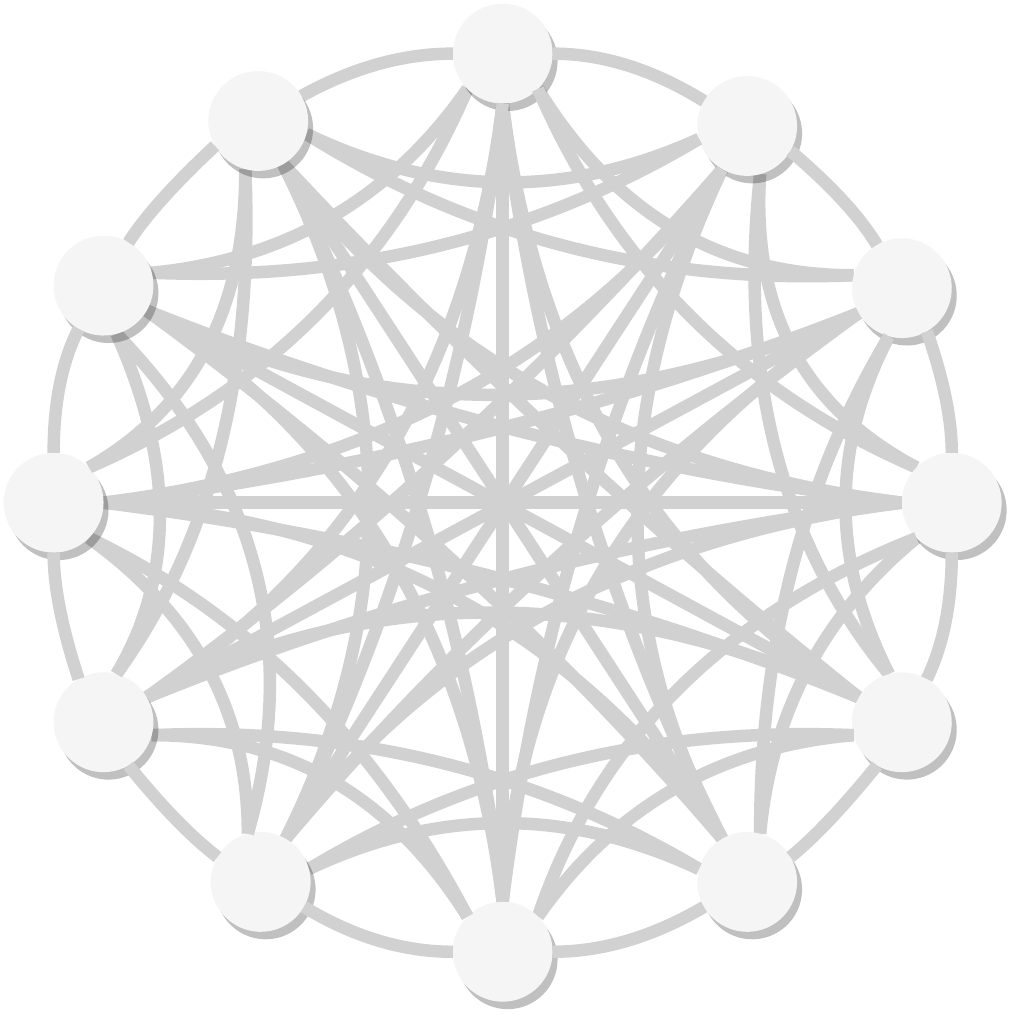}
                \caption{}
                \label{diag:complete}
            \end{center}
        \end{subfigure}
        \vskip 0.25cm
        \caption{Network diagrams with different levels of connectivity. (a) Regular lattice network with $k = 2$. (b) Regular lattice network with $k = 4$.(c) Totally-connected network. }
        \label{diag:network-topologies}
    \end{center}
\end{figure}

We now investigate how applying constraints to the interaction network, i.e.\ limiting the micro-level dynamics, affects the macro-level performance of our model for collective learning.
In the remainder of this paper, we will study interaction networks which take the form of regular lattice networks with different levels of connectivity. A regular lattice network can be defined by its connectivity $k \in [2, m-1]$ where $k$ determines how many ``nearest neighbours'' each agent is connected to when the agents are arranged around a ring. Examples of lattice networks for $k=2$ and $k=4$ are shown in \Cref{diag:swn-k2,diag:swn-k4}, respectively, and are compared with a totally-connected network equivalent to $k = 19$ in \Cref{diag:complete}.

\subsection{Convergence Results for Constrained Interaction Networks}
\label{ssec:interaction-results}

Results are presented for interaction networks with differing levels of connectivity $k$. We continue to vary the connectivity of the physical network via the communication radius to observe whether the physical layer, in conjunction with the interaction layer, has some effect on system performance that wasn't observed previously.

\begin{figure*}[t!]
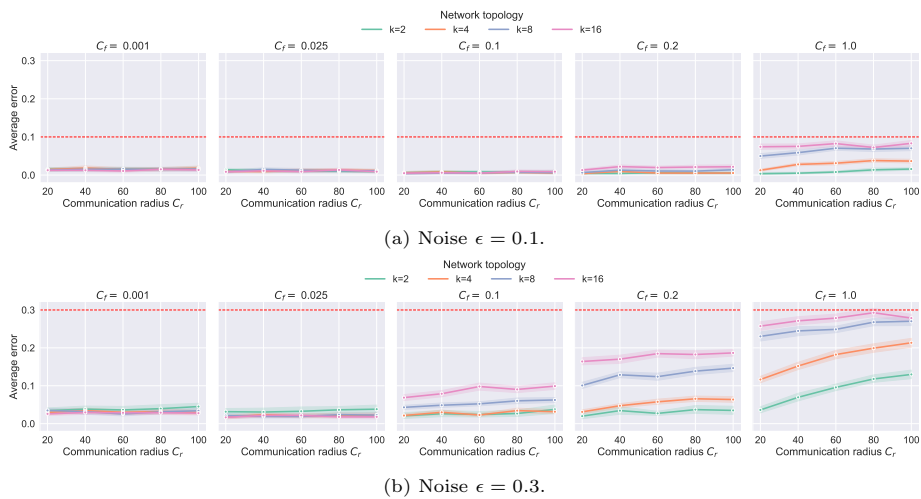

    \begin{center}
        \begin{subfigure}{1\textwidth}
            \includegraphics[width=1\textwidth]{{SWN_avg_error_comms_freq_noise_0.1}.pdf}
            \subcaption[]{Noise $\epsilon = 0.1$.}
            \label{fig:swn-error-0.1}
        \end{subfigure}
        \begin{subfigure}{1\textwidth}
            \includegraphics[width=1\textwidth]{{SWN_avg_error_comms_freq_noise_0.3}.pdf}
            \subcaption[]{Noise $\epsilon = 0.3$.}
            \label{fig:swn-error-0.3}
        \end{subfigure}
        \vskip 0.25cm
        \caption{Average error at steady state against communication radius $C_r$ for increasing communication frequency $C_f = [0, 1]$. Each dotted line shows a different lattice network with connectivity $k \in \{2, 4, 8, 16\}$. The shaded regions show a $95\%$ confidence interval.}
        \label{fig:swn-error-comp}
    \end{center}
\end{figure*}

In \Cref{fig:swn-error-comp} we show a comparison of the average error of the system at steady state against communication radius $C_r$ for different communication frequencies $C_f$. Each solid line represents a different regular lattice network of connectivity $k$. In addition, we compare results for noise $\epsilon = 0.1$ (top) against $\epsilon = 0.3$ (bottom), and show the noise level as a red, dotted line. Beginning with \Cref{fig:swn-error-0.1} we can see that the low-noise scenario converges to an average error at or around $0$ for communication frequencies $C_f < 0.2$ with the best performance observed for $C_f = 0.1$. We also see that the physical connectivity of the system has little impact when comparing across communication radius $C_r$. In the high-noise scenario of \Cref{fig:swn-error-0.3}, however, we see that the average error remains above $0$ for all values of $C_r$ and $C_f$ and that, when $C_f \geq 0.1$, there is a sudden increase in average error for $k \geq 4$. Most importantly, we observe that the networks with the highest connectivity exhibit the largest increases in average error as $C_f$ increases, such that $k = 16$ always results in worse performance than $k = 8$, and so forth. An exception occurs in the high-noise scenario with very low communication frequencies $C_f < 0.1$, where a network with $k=2$ performs slightly worse than more connected networks. However, for $C_f \geq 0.1$, $k=2$ performs on par or better than more connected networks. For example, at the extreme with a communication frequency $C_f = 1$, a network of $k = 2$ has $\frac{1}{5}$ the average error compared with $k = 16$ for the same communication radius $C_r = 20$. When we increase the radius $C_r = 100$, a network of $k = 2$ achieves less than $\frac{1}{2}$ the average error of a network with $k = 16$.
Furthermore, the impact of the physical network also becomes more apparent in the high-noise scenario where, for $C_f = 1$, we see a clear increase in average error as the communication radius $C_r$ is increased.

\begin{figure*}[t!]
    \begin{center}
        \begin{subfigure}{1\textwidth}
            \includegraphics[width=1\textwidth]{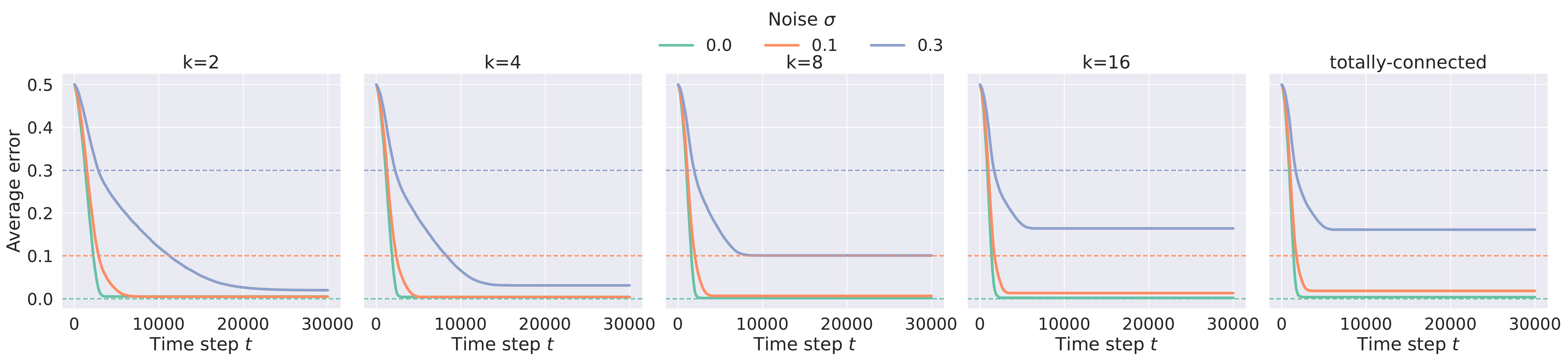}
        \end{subfigure}
        \vskip 0.25cm
        \caption{Average error against time $t$ for $C_r = 20$ and $C_f = 0.2$ for different regular lattice networks with connectivity $k \in \{2, 4, 8, 16\}$ and a totally-connected network (i.e. $k=19$). Each dotted line shows the noise $\epsilon \in \{0, 0.1, 0.3\}$. The shaded regions show a $95\%$ confidence interval.}
        \label{fig:comparison-trajectories}
    \end{center}
\end{figure*}

To better understand the dynamics of our model and how the underlying connectivity affects time to convergence, we present trajectories of the average error against time in \Cref{fig:comparison-trajectories}. Each plot represents a network of different connectivity $k$, and each line shows that network's performance for different noise $\epsilon$. We have fixed $C_r = 20$ and $C_f = 0.2$ as the best performing parameters in this comparison. From left to right, as connectivity increases, we see that for a high-noise scenario the time to convergence is much slower, with the $k=2$ network converging after around $25,000$ time steps. This is reduced to less than $15,000$ time steps for $k=4$ but with slightly higher average error. Then, as we continue to increase the connectivity $k$, the network continues to converge in less time but with greater average error. Therefore, increasing the connectivity worsens performance when the noise is high.

These results suggest that, for higher communication frequencies, the combination of both the interaction network and the physical network is an important consideration.
It is also clear that as the interaction network becomes increasingly connected, the connectivity of the physical network has less of an impact, and that the inverse is true where, for a less connected interaction network, the physical network has greater impact overall. Therefore, both networks are important for considering optimum performance in distributed autonomous systems.

\section{Conclusion and Future Work}
\label{sec:conclusion}

In this paper we have proposed to study the effects of both the physical and interaction networks on the collective learning performance of a multi-agent system. We have demonstrated that, when the underlying interaction network is totally connected, the physical network has minimal impact on performance in our model. However, when the connectivity of the interaction network is greatly reduced the system performs better, achieving lower average error when compared with highly-connected interaction networks under the same experimental conditions. Limiting the interactions between agents also results in improved robustness to noise, where increased connectivity at the physical layer only worsens performance for the same limited interaction network.
Overall, performance is best when the connectivity of both the physical and interaction networks is severely reduced, e.g., for a connectivity of $k=2$ and a communication radius $C_r = 20$.

To our knowledge, this is the first such study explicitly investigating the impact of both the physical and interaction networks in combination in a swarm setting. While it is difficult to draw strong, broad conclusions from this study, we hope that this will encourage other researchers working on collective behaviours to consider the importance of the underlying interaction network and its level of connectivity. Obviously, this study is limited to simulation studies of the effects of the interplay between the physical and interaction network, but a more theoretical study of this interplay is important to understand the reason behind the observed dynamics.

Up to now we have only considered regular lattice networks and small-world networks as the basis for interaction networks. These networks possess certain desirable properties which may be more conducive to the processes involved in collective learning than may other, different, types of networks.
Furthermore, we have only considered the effect of network connectivity in the context of collective learning. This effect may not necessarily be observed for other kinds of collective behaviours. As such, we intend to further explore how network connectivity, and information sharing more generally, impacts different kinds of collective behaviours in multi-agent systems. We are also interested in understanding how a more dynamic environment might change the optimal level of connectivity in the networks, as less connectivity may lead to outdated information being retained by the population which is likely to affect performance over time.

\section{Acknowledgements}
\label{sec:acknowledgements}

This work was funded and delivered in partnership between Thales Group, University of Bristol and with the support of the UK Engineering and Physical Sciences Research Council, ref. EP/R004757/1 entitled ``Thales-Bristol Partnership in Hybrid Autonomous Systems Engineering (T-B PHASE).''

\bibliographystyle{spmpsci}
\bibliography{main}

\end{document}